\documentclass[12pt]{iopart}

\usepackage[dvips]{graphicx}
\usepackage[usenames]{color}

\def\gsim{\mathop {\vtop {\ialign {##\crcr 
$\hfil \displaystyle {>}\hfil $\crcr \noalign {\kern1pt \nointerlineskip } 
$\,\sim$ \crcr \noalign {\kern1pt}}}}\limits}
\def\lsim{\mathop {\vtop {\ialign {##\crcr 
$\hfil \displaystyle {<}\hfil $\crcr \noalign {\kern1pt \nointerlineskip } 
$\,\,\sim$ \crcr \noalign {\kern1pt}}}}\limits}
  
\begin{document}

\title{New Universality Class of Quantum Criticality in Ce- and Yb-based Heavy Fermions}

\author{Shinji Watanabe$^{1}$ and Kazumasa Miyake$^{2}$}

\address{
$^1$ Faculty of Engineering, Kyushu Institute of Technology, Kitakyushu, Fukuoka 804-8550, Japan 
\\
$^2$ Graduate School of Engineering Science, Osaka University, Toyonaka, Osaka 560-8531, Japan}
\ead{}
\begin{abstract}
A new universality class of quantum criticality 
emerging in itinerant electron systems with strong local electron 
correlations is discussed. 
The quantum criticality of a Ce- or Yb-valence transition gives us a unified explanation for 
unconventional criticality 
commonly observed in heavy fermion metals such as YbRh$_2$Si$_2$ and $\beta$-YbAlB$_4$, 
YbCu$_{5-x}$Al$_{x}$, and CeIrIn$_5$.  
The key origin is due to the locality of the critical valence fluctuation mode 
emerging near the quantum critical end point of the first-order valence transition, 
which is caused by strong electron correlations for f electrons.   
Wider relevance of this new criticality and important future measurements to uncover 
its origin are also discussed. 
\end{abstract}

\maketitle

\section{Introduction}

Quantum critical phenomena in strongly correlated electron systems, 
which do not follow conventional quantum critical phenomena for  
spin fluctuations~\cite{Moriya,MT,Hertz,Millis},  
have attracted much attention.  
At low temperatures in the paramagnetic-metal phase in 
YbRh$_2$Si$_2$~\cite{Trovarelli,Gegenwart2007} and 
YbRh$_2$(Si$_{0.95}$Ge$_{0.05}$)$_2$~\cite{Custers,Gegenwart2005},  
uniform magnetic susceptibility is enhanced as $\chi(T)\sim T^{-0.6}$,  
specific heat coefficient shows a logarithmic divergence $C/T\sim -\log{T}$ 
for $0.3~{\rm K} \le T \le 10$~K, 
giving rise to large Wilson ratio $R_{\rm W}\approx 17.5$ 
at $T=90$~mK~\cite{Gegenwart2005} 
much larger than 
the conventional value of $R_{\rm W}\approx 2$ 
which is expected in systems with locally strong correlation. 
In a wide temperature range for $20~{\rm mK} \le T \le 10$~K, $T$-linear resistivity 
appears~\cite{Custers}. 

To understand the unconventional criticality, some theoretical efforts have been made. Local criticality theory has been intensively discussed~\cite{Si,Coleman} and theory of tricritical point has been proposed~\cite{Misawa}.  However, the mechanism and origin of the unconventional criticality do not seem to be fully clarified. Furthermore, new aspects which seem to be related to the origin of the unconventional criticality have been revealed from experimental side. 

Recently, very similar unconventional criticality has been observed 
in the heavy fermion metal $\beta$-YbAlB$_4$~\cite{Nakatsuji,Matsumoto2010,Matsumoto2011,Matsumoto_JPCS2011}.  
Uniform magnetic susceptibility and specific heat show $\chi(T)\sim T^{-0.5}$ 
and $C/T\sim -\log{T}$ at least for $0.3~{\rm K} \le T \le 2$~K, respectively, 
giving rise to the enhanced Wilson ratio 
$R_{\rm W}\sim 25$  at $T=0.4$~K~\cite{Matsumoto2011,Matsumoto_JPCS2011}, and $T$-linear resistivity appears in a wide temperature region  for $1~{\rm K}\le T \le 4$~K~\cite{Nakatsuji}. 
Hard X-ray photoemission measurement has revealed that 
the valence of Yb is +2.75 at $T=20$~K, indicating that $\beta$-YbAlB$_4$ is 
an intermediate valence material, which suggests the importance of Yb-valence 
fluctuations~\cite{Okawa}. 

The isostructural first-order valence transition occurs in the temperature-pressure $(P)$ phase diagram in Ce metal well known as $\gamma$-$\alpha$ transition where the critical end point (CEP) is located at $T_{\rm CEP}\sim 600$~K and $P\sim 2$~GPa~\cite{Ce}. At $x\sim 0.1$ in Ce$_{0.9-x}$Th$_{0.1}$La$_{x}$,
$T_{\rm CEP}$ is suppressed to be close to $T=0$~K. Although $T_{\rm CEP}$ seems to be not exactly zero but finite, at low temperatures $T$-linear resistivity appears and uniform magnetic susceptibility is enhanced. The Wilson ratio $R_{\rm W}\sim 3$ is not extremely large, but is larger than 2~$[18]$, which is expected to be realized in the system with strong local electron correlations. These suggest the importance of critical Ce-valence fluctuations for the unconventional criticality. 

In this paper, we discuss that quantum criticality of Yb- and 
also Ce-valence fluctuation is a key origin of these unconventional criticality.  
We outline a theoretical framework for quantum critical phenomena of the 
valence transition. 
Then, we show that unconventional criticality commonly observed 
not only in YbRh$_2$Si$_2$ and $\beta$-YbAlB$_4$ but also in other materials 
such as YbCu$_{5-x}$Al$_{x}$ and CeIrIn$_5$ 
can be naturally explained by quantum valence criticality. 
We show that fundamental properties of the materials are understood coherently from this viewpoint. 
We discuss that quantum valence criticality offers 
a new class of universality in itinerant electron systems with strong local electron 
correlations.

\section{Theory of quantum critical phenomena of valence fluctuations}

\subsection{Model}

We discuss electronic states of Ce- and Yb-based heavy fermion systems 
on the basis of a generalized periodic Anderson model with inter-orbital Coulomb repulsion 
as the simplest minimal model: 
\begin{eqnarray}
{\cal H}_{\rm PAM}&=&
\sum_{{\bf k}\sigma}\varepsilon_{\bf k}c^{\dagger}_{{\bf k}\sigma}c_{{\bf k}\sigma}
+\varepsilon_{\rm f}\sum_{i\sigma}n^{\rm f}_{\sigma}+\sum_{{\bf k}\sigma}\left(V_{\bf k}
f^{\dagger}_{{\bf k}\sigma}c_{{\bf k}\sigma}+{\rm H.C.}\right)
+U\sum_{i}n_{i{\uparrow}}^{\rm f}n_{i{\downarrow}}^{\rm f}
\nonumber
\\
&+&U_{\rm fc}\sum_{i\sigma\sigma'}n^{\rm f}_{i\sigma}n^{\rm c}_{i\sigma'}, 
\label{eq:PAM}
\end{eqnarray}
where $f_{i\sigma}$ and $c_{i\sigma}$ are the annihilation operator of the f electron   
and the conduction electron at the $i$-th site with a spin $\sigma$, respectively. 
Here, $n^{\rm a}_{i\sigma}$ $({\rm a}={\rm f}, {\rm c})$ are defined by  
$n^{\rm f}_{i\sigma}\equiv f^{\dagger}_{i\sigma}f_{i\sigma}$ and 
$n^{\rm c}_{i\sigma}\equiv c^{\dagger}_{i\sigma}c_{i\sigma}$, respectively. 
$\varepsilon_{\bf k}$ is the energy band of conduction electrons. 
$\varepsilon_{\rm f}$ is the f level. 
$V_{\bf k}$ is the c-f hybridization. 
$U$ is the onsite Coulomb repulsion between f electrons. 
$U_{\rm fc}$ is the inter-orbital Coulomb repulsion, which is considered to be 
important in causing the first-order valence transition (FOVT). 
Since Ce$^{3+}$ is a 4f$^1$ configuration and Ce$^{4+}$ is a 4f$^0$ configuration, 
eq.~(\ref{eq:PAM}) describes the electronic states for Ce-based heavy fermion systems. 
On the other hand, 
since Yb$^{3+}$ is a 4f$^{13}$ configuration and Yb$^{2+}$ is 
a 4f$^{14}$ configuration, which is a closed shell, 
if we take the hole picture instead of the electron picture, 
eq.~(\ref{eq:PAM}) describes electronic states for Yb-based heavy fermion systems. 

The model Hamiltonian (\ref{eq:PAM}) may be regarded as a generalization of the 
Falicov-Kimball model~\cite{Falicov} in which the c-f hybridization is missing, and 
also a specified version of those discussed by Varma at early stage of 
research on valence fluctuation phenomena~\cite{Varma1976}. 
A variety of previous works based on Hamiltonian (\ref{eq:PAM}) were briefly 
reviewed in refs.~\cite{OnishiM, Miyake2007}.  
Similar effects of the d-p Coulomb interaction in the so-called 
d-p model have been investigated as a possible charge-transfer 
fluctuation mechanism of the high $T_{\rm c}$ superconductor~\cite{Varma1987,Hirashima}. 

\subsection{Phase diagram}

\begin{figure}
\begin{center}
\includegraphics[width=85mm]{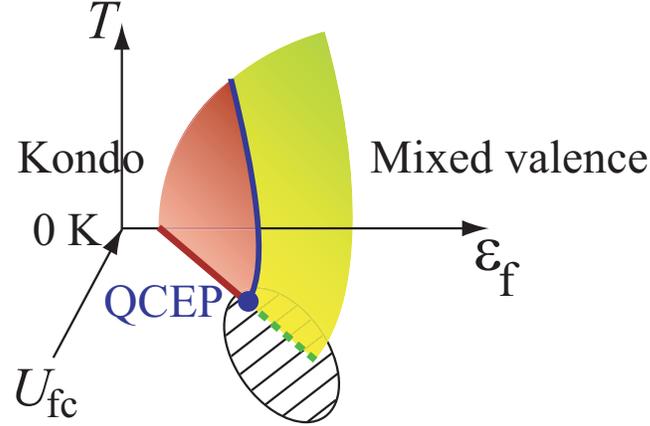}
\end{center}
\caption{\label{fig:TEfUfc}(color online) 
$T$-$\varepsilon_{\rm f}$-$U_{\rm fc}$ phase diagram of $\cal H_{\rm PAM}$. 
The first-order valence transition surface (dark surface) extends to 
the valence-crossover surface (light surface). 
The critical end line (solid line) touches $T=0$~K at which quantum critical end point 
(QCEP) is realized. Superconductivity emerges near QCEP as illustrated by the hatched area 
(see text). 
Note that the location of the QCEP depends on the strength of the c-f hybridization in (1). When the c-f hybridization becomes smaller, the QCEP shifts to the smaller $U_{\rm fc}$ side. 
}
\end{figure}

To clarify the fundamental properties of the model ${\cal H}_{\rm PAM}$, 
intensive efforts have been made:  
Mean-field theories were applied to ${\cal H}_{\rm PAM}$  
in the $d=3$~\cite{Watanabe2009}, $d=2$~\cite{WM2010}, and 
$d=1$~\cite{Watanabe2006} systems with $d$ being a spatial dimension. 
Numerical calculations by   
the DMRG in the $d=1$ system~\cite{Watanabe2006} and 
the DMFT in the $d=\infty$ system~\cite{Saiga2008} 
were also performed. 
The phase diagram of ${\cal H}_{\rm PAM}$ 
for a typical set of parameters of Ce- and Yb-based heavy fermion systems 
is shown schematically in Fig.~\ref{fig:TEfUfc}. 
The FOVT surface (dark surface) extends to 
the sharp valence crossover surface (light surface). 
The critical end line is formed between them, which is the edge of 
the first-order transition surface. 
The quantum critical end point (QCEP) of the FOVT is the point 
at which the critical end line touches the $\varepsilon_{\rm f}$-$U_{\rm fc}$ 
plane of the ground state. 
At the critical end line as well as the QCEP, valence fluctuation susceptibility 
\begin{eqnarray}
\chi_{\rm v}=-\frac{\partial n_{\rm f}}{\partial\varepsilon_{\rm f}}
\label{eq:VF}
\end{eqnarray}
diverges, i.e., $\chi_{\rm v}=\infty$. 
Here, $n_{\rm f}\equiv\sum_{{i}\sigma}\langle n_{i\sigma}^{\rm f}\rangle/N_{\rm s}$ 
with $N_{\rm s}$ being the number of sites.  
In the small-$\varepsilon_{\rm f}$ limit, f electron number per site 
$n_{\rm f}$ becomes $n_{\rm f}=1$, which is called the Kondo regime, 
corresponding to Ce$^{+3}$ (4f$^1$ electron per site) or Yb$^{+3}$ 
(4f$^1$ hole per site) state.
As $\varepsilon_{\rm f}$ increases, $n_{\rm f}$ decreases, and  
at the FOVT surface, $n_{\rm f}$ shows a discontinuous decrease,  
while at the valence crossover surface, $n_{\rm f}$ shows a continuous but sharp decrease 
with enhanced valence fluctuation. 
When $\varepsilon_{\rm f}$ further decreases, $n_{\rm f}$ further decreases, 
which is called mixed the valence regime and corresponds to the 
Ce$^{+3+\delta}$ or Yb$^{+3-\delta}$ state ($\delta$ is a positive number smaller than 1) 
i.e., intermediate valence state. 
Note that $n_{\rm f}$ at the FOVT surface and valence-crossover surface depends on 
the details of model parameters in eq.~(\ref{eq:PAM}) such as momentum dependence of 
c-f hybridization $V_{\bf k}$ and band structures $\varepsilon_{\bf k}$. 
Namely, at the FOVT as well as valence-crossover surface, 
$n_{\rm f}$ can be close to 1 and also can be $1-\delta$. 
Hence, note that the terms ``Kondo" and ``mixed valence" in Fig.~\ref{fig:TEfUfc} merely represent a relatively larger-$n_{\rm f}$ state 
and relatively smaller-$n_{\rm f}$ state. 

The critical valence fluctuation arising from the QCEP as well as 
the critical end line in Fig.~\ref{fig:TEfUfc} causes several anomalies: 
Valence fluctuation mediated superconductivity appears 
in the Kondo regime near the QCEP~\cite{OnishiM,Watanabe2006}  
as shown by the hatched area at $T=0$~K in Fig.~\ref{fig:TEfUfc}. 
Near the QCEP,  
residual resistivity is enhanced~\cite{MM} 
and $T$-linear resistivity appears~\cite{HJM}. 
Heavy fermion metals 
CeCu$_2$Ge$_2$~\cite{Jaccard}, CeCu$_2$Si$_2$~\cite{Holms}, and  
CeCu$_2$(Ge$_x$Si$_{1-x}$)$_2$~\cite{Yuan}, 
where these anomalies were observed, 
are considered to be located in the valence crossover regime 
with the superconducting phase in Fig.~\ref{fig:TEfUfc}. 
Since inter-orbital repulsion $U_{\rm fc}$ is considered to have 
the inter-site character in Ce- and Yb-based compounds, most of the compounds  
seem to have moderate values of $U_{\rm fc}$, which are located 
in the valence crossover regime in Fig.~\ref{fig:TEfUfc}. 
For details, readers can refer to refs.~\cite{Miyake2007,wataPSS2010}. 

Since the critical end line extends from the QCEP to the negative temperature region 
in Fig.~\ref{fig:TEfUfc}, the CEP can be regarded to be located at negative temperatures for $U_{\rm fc}<U_{\rm fc}^{\rm QCEP}$ with $U_{\rm fc}^{\rm QCEP}$ being $U_{\rm fc}$ at the QCEP. Namely, the valence crossover line for $T \ge 0$ (on the light surface 
in Fig.~\ref{fig:TEfUfc}) 
is extended from the virtual CEP of the FOVT at negative temperatures 
(see Fig.~\ref{fig:TPD}(c)). 
This CEP at negative temperature is an analogous concept for the negative Curie or N{\'e}el temperature of nearly ferromagnetic or antiferromagnetic materials. 
We note that in the valence crossover regime in Fig.~\ref{fig:TEfUfc}, the CEP at negative temperatures is moved up toward the $T>0$ direction by applying a magnetic field, giving rise to an emergence of the QCEP~\cite{Watanabe2011}. Namely, it has been shown that the CEP is induced to emerge from the negative temperature side to the positive temperature side by applying a magnetic field in the valence-crossover regime~\cite{wataPSS2010,WTMF2008}.
Correspondence between the phase diagram in Fig.~\ref{fig:TEfUfc} 
and the $T$-$P$ or $T$-$H$ phase diagram in Ce- and Yb-based systems 
is explained in detail in ref.~\cite{Watanabe2011}.

\subsection{Quantum valence criticality}

In model (\ref{eq:PAM}), the on-site Coulomb repulsion $U$ 
for f electrons is the strongest interaction. 
Therefore, we should first take into account 
the effects of the $U$ term. 
Then, the mode-coupling theory for critical valence fluctuations 
caused by the $U_{\rm fc}$ term should be constructed.  
To construct such a theoretical framework, we employ the large-$N$ expansion 
scheme. 
Namely, $\sigma=\uparrow,\downarrow$ is generalized to $m=1\cdots N$ 
in eq.~(\ref{eq:PAM}). 
By imposing a constraint on f-electron number by introducing the Lagrange multiplier 
$\lambda_{i}$ as 
$-\sum_{i}\lambda_{i}\left(\sum_{m}f^{\dagger}_{im}f_{im}+Nb^{\dagger}_{i}b_{i}-1\right)$
with $b_{i}$ being a slave-boson operator, 
we first take the $U=\infty$ limit in eq.~(\ref{eq:PAM}). 
Based on the saddle-point solution for $U=\infty$, 
the action of the system is obtained as a perturbation expansion with respect to 
the $U_{\rm fc}$ term 
by introducing the identity applied by 
a Stratonovich-Hubbard transformation
$
{\rm e}^{-S'}
=\int{\cal D}\varphi
\exp[
\sum_{im}
\int_{0}^{\beta}d\tau
\{
-\frac{U_{\rm fc}}{2}\varphi_{im}(\tau)^2
+i\frac{U_{\rm fc}}{\sqrt{N}}
(
c_{im}f^{\dagger}_{im}-f_{im}c^{\dagger}_{im}
)
\varphi_{im}(\tau)
\}
]
$ 
to $S'=\int_{0}^{\beta}d\tau{\cal L'}(\tau)$ with 
${\cal L'}=-U_{\rm fc}\sum_{im}(n_{im}^{\rm c}+n_{im}^{\rm f})/2
+U_{\rm fc}\sum_{imm'}n_{im}^{\rm f}n_{im'}^{\rm c}/N$.  
Namely, the action $S=S_{0}+S'$ where $S_{0}$ is the action for the terms without 
$U_{\rm fc}$ in (\ref{eq:PAM}) is given by 
\begin{eqnarray}
& &S\left[\varphi\right]=\sum_{m}\left[
\frac{1}{2}
\sum_{\bar{q}}\Omega_{2}(\bar{q})
\varphi_{m}(\bar{q})\varphi_{m}(-\bar{q})
\right.
\nonumber
\\
&+&
\left.
\sum_{\bar{q}_1,\bar{q}_2,\bar{q}_3}
\Omega_{3}(\bar{q}_1,\bar{q}_2,\bar{q}_3)
\varphi_{m}(\bar{q}_1)
\varphi_{m}(\bar{q}_2)
\varphi_{m}(\bar{q}_3)
\delta\left(\sum_{i=1}^{3}\bar{q}_{i}\right)
\right.
\nonumber
\\
&+&
\left. 
\sum_{\bar{q}_1,\bar{q}_2,\bar{q}_3,\bar{q}_4}
\Omega_{4}(\bar{q}_1,\bar{q}_2,\bar{q}_3,\bar{q}_4)
\varphi_{m}(\bar{q}_1)
\varphi_{m}(\bar{q}_2)
\varphi_{m}(\bar{q}_3)
\varphi_{m}(\bar{q}_4)
\delta\left(\sum_{i=1}^{4}\bar{q}_{i}\right)
+\cdots
\right]
\label{eq:Fexpand}
\end{eqnarray}
where $\Omega_{2}(\bar{q})\approx U_{\rm fc}
\left[1-2U_{\rm fc}\chi_{0}^{\rm ffcc}(\bar{q})/N\right]$, 
$\Omega_{3}(\bar{q}_1,\bar{q}_2,\bar{q}_3)=O(U_{\rm fc}^3)$, 
and 
$\Omega_{4}(\bar{q}_1,\bar{q}_2,\bar{q}_3,\bar{q}_4)=O(U_{\rm fc}^4)$ 
with $\bar{q}\equiv({\bf q},i\omega_l)$ and $\omega_{l}=2l{\pi}T$. 
Here, $\chi_{0}^{\rm ffcc}(\bar{q})$ is given by 
$\chi_{0}^{\rm ffcc}(\bar{q})=-T\sum_{{\bf k},n}
G^{\rm ff}_{0}({\bf k}+{\bf q},i\varepsilon_{n}+i\omega_{l})
G^{\rm cc}_{0}({\bf k},i\varepsilon_{n})/N_{\rm s}$ 
with  $\varepsilon_{n}=(2n+1){\pi}T$. 
Here, $G_{0}^{\rm ff}({\bf k},i\varepsilon_{n})$ and 
$G_{0}^{\rm cc}({\bf k},i\varepsilon_{n})$ are the Green functions 
of f and conduction electrons for the saddle point solution for 
$U=\infty$, respectively~\cite{VQCP2010}, and 
$N_{\rm s}$ is the number of lattice sites. 

Since long wavelength $|{\bf q}|\ll q_{\rm c}$ around ${\bf q}={\bf 0}$ 
and low frequency $|\omega|\ll\omega_{\rm c}$ regions play dominant role 
in critical phenomena with $q_{\rm c}$ and $\omega_{\rm c}$ being cutoffs 
for for momentum and frequency, respectively, $\Omega_{i}$ are expanded 
for $q$ and $\omega$ around $({\bf 0},0)$ as 
$\Omega_{2}(\bar{q})\approx \eta+Aq^2+C_{q}|\omega_{l}|$, 
$\Omega_{3}(\bar{q}_1,\bar{q}_2,\bar{q}_3)\approx v_{3}\sqrt{T/N_{\rm s}}$, and 
$\Omega_{4}(\bar{q}_1,\bar{q}_2,\bar{q}_3,\bar{q}_4)\approx v_{4}T/N_{\rm s}$. 
Let us here apply 
the Hertz's renormalization-group procedure~\cite{Hertz} to $S[\varphi]$: 
(a) Integrating out high momentum and frequency parts for $q_{\rm c}/s<q<q_{\rm c}$ and 
$\omega_{\rm c}/s^{z}<\omega<\omega_{\rm c}$, respectively, with 
$s$ being a dimensionless scaling parameter $(s\ge 1)$ and $z$ the dynamical exponent. 
(b) Scaling of $q$ and $\omega$ by $q'=sq$ and $\omega'=s^{z}\omega$. 
(c) Re-scaling of $\varphi$ by $\varphi'({\bf q'},\omega')=s^{a}\varphi({\bf q'}/s,\omega'/s)$. 
Then, we dertermined that to make the Gaussian term in Eq.~(\ref{eq:Fexpand}) scale invariant, 
$a$ must satisfy $a=-(d+z+2)/2$ with $d$ spatial dimension and the dynamical exponent $z=3$. 
The renormalization-group equations for coupling constants $v_{j}$ are derived as 
$
\frac{dv_{3}}{ds}=\left[6-(d+z)\right]v_{3}+O(v_{3}^2), 
$
and 
$
\frac{dv_{4}}{ds}=\left[4-(d+z)\right]v_{4}+O(v_{4}^2),
$
for cubic and quadratic terms, respectively. 
By solving these equations, it is shown that higher order terms than 
the Gaussian term are irrelevant 
\begin{eqnarray}
\lim_{s\to\infty}v_{j}(s)=0 \ \ {\rm for} \ \ j\ge 3
\label{eq:vterms}
\end{eqnarray}
for $d+z>6$. 
For the case of $d=3$ and $z=3$, it is shown
that the cubic term is marginally irrelevant~\cite{Miyake2007}. 
It is noted that 
in $d=3$ systems for realistic heavy fermion systems, 
eq~(\ref{eq:vterms}) is considered to hold as discussed in the end of section~2.3.   

We have found that 
almost dispersionless critical valence fluctuation modes appear 
near $q=0$ not only for deep $\varepsilon_{\rm f}$, i.e., in the Kondo regime, 
but also for shallow $\varepsilon_{\rm f}$, i.e., in the mixed valence regime,  
because of strong on-site Coulomb repulsion for f electrons 
in eq.~(\ref{eq:PAM})~\cite{VQCP2010}. 
This ``almost local" nature causes the extremely small coefficient $A$ in 
dynamical valence susceptibility defined by the inverse of the coefficient 
of the Gaussian term $\Omega_{2}(\bar{q})$ in the action 
\begin{eqnarray}
\chi_{\rm v}(q,i\omega_{l})=\frac{1}{\eta+Aq^2+C_{q}|\omega_{l}|}, 
\label{eq:chiv}
\end{eqnarray}
where $C_{q}=C/{\rm max}\{q,l_{\rm i}^{-1}\}$ with $l_{\rm i}$ being 
the mean free path of impurity scattering~\cite{Miyake1999} and 
$\omega_{l}$ is the Boson Matsubara frequency $\omega_{l}=2\pi lT$. 
This yields $C_{q}=C/q$ in the clean system and $C_{q}=C/l_{\rm i}$ 
in the dirty system affected by impurity scattering,
giving rise to the dynamical exponent $z=3$ and $z=2$, respectively. 

The emergence of the weak-$q$ dependence in the critical valence fluctuation is analyzed, as follows~\cite{Miyake2007}: Since $\chi_{\rm v}(\bar{q})\equiv\Omega_{2}^{-1}(\bar{q})$, the $q$-dependence in eq.~(\ref{eq:chiv}) appears through $G^{\rm ff}_{0}(\bar{k}+\bar{q})$ in $\chi_{0}^{\rm ffcc}(\bar{q})$. Near $q=0$, $\chi^{\rm ffcc}_{0}(q,0)$ is expanded as
\begin{eqnarray}
\hspace*{2cm}
\chi_{0}^{\rm ffcc}(q,0)=\chi_{0}^{\rm ffcc}(0,0)
+\tilde{S}\left(\frac{V}{|\mu-\varepsilon_{\rm f}|}\right)^2q^2,
\label{eq:chi_0_ffcc}
\end{eqnarray}
%
where $\tilde{S}$ includes the effect of the f-electron self-energy for $U$ in eq.~(\ref{eq:PAM}). Since the f-electron self-energy has almost no $q$ dependence in heavy electron systems, the $q$  dependence of the f-electron propagator $G_{0}^{\rm ff}(\bar{k}+\bar{q})$ comes from the hybridization $V$ with conduction electrons with the dispersion $\varepsilon_{{\bf k}+{\bf q}}$, as seen in the coefficient of the $q^2$ term in eq.~(\ref{eq:chi_0_ffcc}).  Hence, the reduction of the coefficient $A$ in eq.~(\ref{eq:chiv}) is caused by two factors. One is due to the smallness of $(V/|\mu-\varepsilon_{\rm f}|)^2$. In typical heavy electron systems, this factor is smaller than $10^{-1}$. The other is the reduction of the coefficient $\tilde{S}$, which is suppressed by the effects of the on-site electron correlations $U$ in eq.~(\ref{eq:PAM}). Numerical evaluations of $\chi_{0}^{\rm ffcc}(q,0)$ based on the saddle point solution for $U=\infty$ in eq.~(\ref{eq:PAM}) show that extremely small $\tilde{S}$ appears not only in the Kondo regime, but also in the mixed-valence regime~\cite{VQCP2010}, indicating that the reduction by $\tilde{S}$ plays a major role. These multiple reductions are the reason why extremely small coefficient $A$ appears in eq.~(\ref{eq:chiv}). 

The extremely small $A$ in eq.~(\ref{eq:chiv}) 
makes the characteristic temperature for critical valence fluctuations
\begin{eqnarray}
T_{0}\equiv\frac{Aq_{\rm B}^3}{2\pi C}
\label{eq:T0}
\end{eqnarray}
extremely small. 
Here, $q_{\rm B}$ is a momentum at the Brillouin zone boundary. 
Hence, even at low enough temperature, lower than the effective Fermi temperature 
of the system, i.e., the so-called Kondo temperature, $T\ll T_{\rm K}$, 
the temperature scaled by $T_{0}$ can be very large: $t\equiv T/T_{0}\gg 1$. 
This is the main reason why unconventional criticality emerges at ``low" temperatures, 
which will be explained below. 

By optimizing the action (\ref{eq:Fexpand}) derived from 
${\cal H}_{\rm PAM}$, taking account of the mode coupling effect 
for critical valence fluctuations, we obtain a self-consistent renormalization 
(SCR) equation for critical valence fluctuations: 
\begin{eqnarray}
y=y_{0}+\frac{3}{2}y_{1}t\left[
\frac{x_{\rm c}^3}{6y}-\frac{1}{2y}\int_{0}^{x_{\rm c}}dx\frac{x^3}{x+\frac{t}{6y}}
\right], 
\label{eq:SCReq}
\end{eqnarray}
where 
$y\equiv \eta/(Aq_{\rm B}^2)$, $x\equiv q/q_{\rm B}$, 
$x_{\rm c}\equiv q_{\rm c}/q_{\rm B}$ with $q_{\rm c}$ being a momentum cutoff, 
and 
$y_0$ parameterizes a distance from the criticality and 
$y_1$ is a dimensionless mode-coupling constant of $O(1)$.  
The solution of 
Eq.~(\ref{eq:SCReq}) is quite different from that of ordinary SCR equations 
for spin fluctutions~\cite{Moriya} 
because of the extremely small $A$ in eq.~(\ref{eq:chiv}). 

In the $y\gg t$ limit, the analytic expression of $\chi_{\rm v}(0,0)$ can be obtained, 
which is $\chi_{\rm v}(0,0)=y^{-1}\sim t^{-2/3}$ for both the clean $(z=3)$ system and 
dirty $(z=2)$ system. 
As shown in ref.~\cite{WTMF2008},  
uniform magnetic susceptibility diverges at the QCEP where valence fluctuation diverges. 
This is due to the fact that $\chi_{\rm v}({\bf q},i\omega_{l})$ and the dynamical f-spin susceptibility $\chi({\bf q},i\omega_{l})=\int_{0}^{\beta}d\tau\langle T_{\tau}S^{+}_{\rm f}({\bf q},\tau)S^{-}_{\rm f}(-{\bf q},\tau) \rangle e^{i\omega_{l}\tau}$ have the common structure near the QCEP~[35]. 
The spin-lattice relaxation rate is given by $(T_{1}T)^{-1}=2(\gamma_{\rm n}^2/g_{\rm f}\mu_{\rm B})^2\sum_{\bf q}|D_{\bf q}|^2{\rm Im}\chi^{\rm R}({\bf q},\omega_{0})/\omega_{0}$, where $\chi^{\rm R}({\bf q},\omega_{0})$ is the retarded dynamical f-spin susceptibility 
with the nuclear resonance frequency $\omega_{0}$, 
$\gamma_{\rm n}$ is the gyromagnetic ratio of nuclear spin, $g_{\rm f}$ is a Lande's g factor for f electrons, and $D_{\bf q}$ is the hyperfine-coupling constant. Then, we have $(T_{1}T)^{-1}\approx  y^{-1}$ near the QCEP. Namely, the uniform magnetic susceptibility and spin-lattice relaxation rate are shown to have the same temperature dependence: $\chi(t)\sim t^{-2/3}$ and $(T_{1}T)^{-1}\sim t^{-2/3}$ for $y\gg t$. 
%
When $T$ is decreased down to $T\sim T_{0}$, $y$ 
in eq.~(\ref{eq:SCReq}) is evaluated as $y\sim t^{0.5}$ by the least square fit of the 
numerical solution of eq.~(\ref{eq:SCReq}). 
Hence, depending on the flatness of critical valence fluctuation mode and 
measured temperature range, $\chi(T)\sim t^{-\alpha}$ and $(T_1T)^{-1}\sim t^{-\alpha}$ 
with $0.5\lsim \alpha \lsim 0.7$ are observed. 

\begin{table}
\caption{\label{blobs}Resistivity, specific-heat coefficient, uniform magnetic susceptibility, 
and spin-lattice relaxation rate in the vicinity of the QCEP of first-order valence transition. 
Exponent $\alpha$ takes the value for $0.5\lsim\alpha\lsim 0.7$ depending on the 
flatness of the critical valence fluctuation mode and measured $T$ range (see text).}
\begin{indented}
\item[]\begin{tabular}{cccc}
\br
$\rho$&$C/T$&$\chi$&$(T_{1}T)^{-1}$\\
\mr
$T$&$-\log{T}$&$T^{-\alpha}$&$T^{-\alpha}$\\
\br
\end{tabular}
\end{indented}
\label{tb:VQCP}
\end{table}

Calculating resistivity 
$\rho(T)$ as 
$
\rho(T)\propto\frac{1}{T}\int_{-\infty}^{\infty}d\omega
\omega n(\omega)[n(\omega)+1]\int_{0}^{q_{\rm c}}dqq^3{\rm Im}\chi_{\rm v}^{\rm R}(q,\omega)
$
with $n(\omega)=1/({\rm e}^{\beta\omega}-1)$ being the Bose distribution function, and 
$
\chi_{\rm v}^{\rm R}(q,\omega)=(\eta+Aq^2-iC_{q}\omega)^{-1}
$, 
a retarded valence susceptibility, we obtain $\rho(t)\propto t$ for the $y\gsim t$ region.
Here, emergence of $T$-linear resistivity comes from the high-temperature limit of 
the Bose distribution function, indicating that the system is described as if it is 
in the classical regime, because the system is in the high-$T$ regime 
in the scaled temperature $t\equiv T/T_{0}\gg 1$, in spite of $T\ll T_{\rm K}$.  
Hence, this mechanism is essentially the same as the emergence of $T$-linear resistivity 
in the electron-phonon system where $\rho(T)\sim T$ appears for $T\gsim \Theta_{\rm D}/5$ 
with $\Theta_{\rm D}$ being Debye temperature~\cite{Abrikosov}. 
This result that the almost flat dispersion of the critical valence fluctuation mode causes 
$T$-linear resistivity implies that dynamical exponent $z$ can be regarded 
as if $z=\infty$. 

%
An evaluation of the valence-fluctuation exchange process in the self-energy by using eq.~(3) gives logarithmic energy dependence in the real part of the self-energy~\cite{WM2006}. A numerical solution of the self-energy gives logarithmic temperature dependence in the specific-heat coefficient $C/t\sim -{\log}t$ even for the temperature regime for $T>T_{0}$ where the uniform magnetic susceptibility and the resistivity behave as $\chi(t)\sim t^{-\alpha}$ with $0.5\lsim \alpha \lsim 0.7$ and $\rho(t)\sim t$, respectively~[36]. Anomalous temperature dependence of the specific-heat coefficient and the resistivity of our theory shares an aspect similar to that based on the marginal Fermi liquid (MFL) assumption for high-$T_{\rm c}$ cuprates~\cite{Varma1996,Varma2002}, 
in which the self-energy is given by $\Sigma(\varepsilon)\propto \varepsilon\left(\ln\varepsilon-{\rm i}|\varepsilon|\right)$. 
Since uniform magnetic susceptibility $\chi(T)$ shows a power-law divergence, 
the Wilson ratio $R_{\rm W}$, i.e., $\chi/(C/T)$, is significantly enhanced: $R_{\rm W}\gg 2$. 
When the experimentally accessible lowest temperature is larger than $T_{0}$, 
the unconventional criticality 
dominates all the physical quantities down to 
the lowest temperature. 
Namely, near the QCEP in Fig~\ref{fig:TEfUfc}, 
unconventional criticality summarized in Table~\ref{tb:VQCP} appears.  

When the experimentally accessible lowest temperature is smaller than $T_{0}$, 
$y\propto t^{4/3}$ $(y\propto t^{3/2})$ is realized in the clean (dirty) system 
for the $t=T/T_{0}\ll 1$ region. 
Hence, resistivity behaves as $\rho\sim t^{5/3}$ in the clean system and 
$\rho\sim t^{3/2}$ in the dirty system in the $t \to 0$ limit 
$(T \ll T_{0}\ll T_{\rm K})$. 
 
We note that the present formulation reproduces the previous results 
for the local limit of critical valence fluctuation mode $A\to 0$ in eq.~(\ref{eq:chiv}), 
which was derived starting from phenomenologically introduced 
dynamical valence susceptibility assuming $z=\infty$~\cite{HJM}.

\begin{figure}
\begin{center}
\includegraphics[width=150mm]{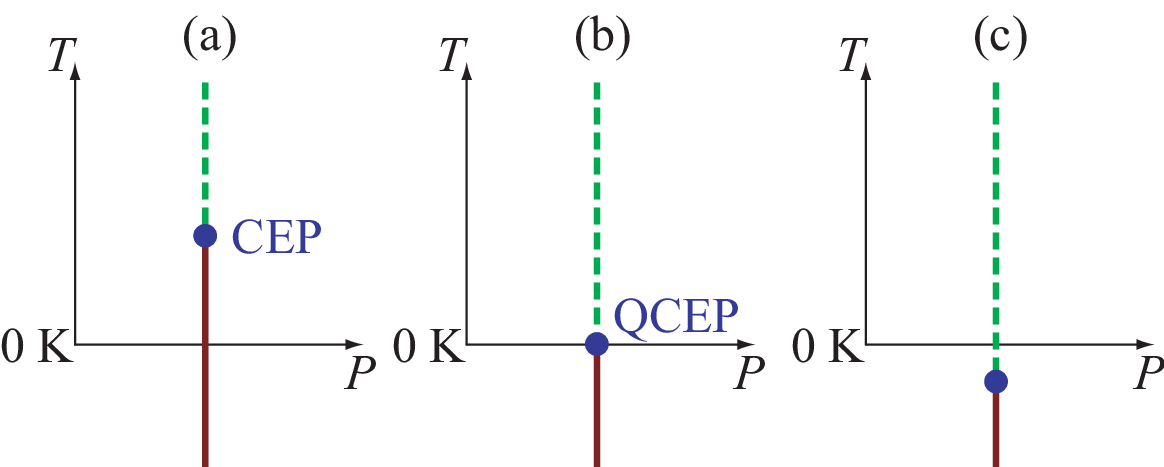}
\end{center}
\caption{(color online) 
Focus on the vicinity of $T=0$~K in $T$-$P$ phase diagram with 
first order valence transition (solid line) terminating at critical end point (CEP) 
(filled circle) from which sharp valence crossover extends (dashed line).
(a) $T_{\rm CEP}>0$ case. (b) $T_{\rm CEP}=0$ case, i.e, 
quantum critical end point (QCEP) is realized. (c) $T_{\rm CEP}<0$ case. 
Note that in Ce (Yb)-based systems, the Kondo regime and mixed-valence regime 
correspond to the smaller (larger) $P$ regime and larger (smaller) $P$ regime, 
respectively. 
This $T$-$P$ phase diagram corresponds to the cutout of the $T$-$\varepsilon_{\rm f}$ plane in Fig.~1 at moderate-$U_{\rm fc}$ region, where the CEP is located at negative temperature (see ref.~\cite{Watanabe2011} for detail).
}
\label{fig:TPD}
\end{figure}

By the third law of thermodynamics, entropy of the system should be zero 
at the ground state. 
This requires that the FOVT line appears in the direction parallel to 
the temperature axis of the phase diagram. 
For example, in the case of the $T$-pressure $(P)$ phase diagram, the slope of the 
FOVT temperature $T_{\rm v}$ is given by 
\begin{eqnarray}
\frac{dT_{\rm v}}{dP}=\frac{V_{\rm K}-V_{\rm MV}}{S_{\rm K}-S_{\rm MV}}
\label{eq:TvH}
\end{eqnarray}
by the Claudius-Clapeyron relation. 
Here, $V$ and $S$ denote the volume and entropy, respectively. 
The subscript K and MV are abbreviations for the Kondo and mixed-valence regimes, 
respectively,   
which represent the relatively larger-$n_{\rm f}$ state and smaller-$n_{\rm f}$ state, 
respectively, as remarked in \S2.2. 
By the third law of thermodynamics, the entropy should be zero at $T=0$~K. 
Hence, the denominator of eq.~(\ref{eq:TvH}) should be zero at the ground state, 
giving rise to $dT_{\rm v}/dP|_{T=0}=\infty$. 
This indicates that the FOVT line should appear 
perpendicularly to the $P$ axis, as shown in Fig.~\ref{fig:TPD}(a). 
Here, the valence crossover line extends from the critical end point 
of the FOVT line.
Hence, the valence-crossover line extended from the QCEP  
appears perpendicularly to the $P$ axis, as shown in Fig.~\ref{fig:TPD}(b). 
Then, when $T$ is decreased to approach the QCEP, 
we usually follow 
the valence crossover line shown by the dashed line in Fig.~\ref{fig:TPD}(b) 
at least in the vicinity of the QCEP.

At the valence crossover line, the cubic term in the action (\ref{eq:Fexpand}) derived from 
${\cal H}_{\rm PAM}$ eq.~(\ref{eq:PAM})~\cite{VQCP2010} 
in the vicinity of the QCEP vanishes, 
which makes the upper critical dimension $d_{\rm u}=4$ of the system, 
but not $d_{\rm u}=6$. 
Then, the clean $(z=3)$ system and the dirty $(z=2)$ system in three spatial dimension $(d=3)$  
are both above the upper critical dimension, i.e., $d+z>d_{\rm u}$. 
Then, the higher order terms other than the Gaussian term are irrelevant in the action, 
which makes the fixed point Gaussian. 
This guarantees the validity of eq.~(\ref{eq:SCReq}) derived by constructing the 
best Gaussian in the action for ${\cal H}_{\rm PAM}$.   
Note that this conclusion holds even when $P$ is replaced by the other 
control parameters such as a magnetic field $H$ in Fig.~\ref{fig:TPD}.

\subsection{Valence fluctuation mediated superconductivity}

Although almost flat dispersion appears near $q=0$ for critical valence fluctuation mode 
as presented above, 
$q$ dependence of the critical mode, i.e., 
a gradual decrease, appears 
around $q=2k_{\rm F}$ with $k_{\rm F}$ being the Fermi wavenumber. 
This moderate $q$ dependence of critical valence fluctuation is shown to 
induce attractive interaction at nearest-neighbor distance in the real space 
by the Fourie transformation~\cite{Miyake2007}. 
Since strong Coulomb repulsion $U$ in eq.~(\ref{eq:PAM}) forbids on-site pairing, 
d-wave pairing is induced by critical valence fluctuations for the spin-singlet sector. 
Indeed, d-wave pairing is shown to appear in the Kondo regime near the QCEP 
in ${\cal H}_{\rm PAM}$ as shown in Fig.~\ref{fig:TEfUfc} 
by the slave-boson mean-field theory taking account of 
Gaussian fluctuations in the $d=3$ system~\cite{OnishiM} 
and the DMRG calculation in the $d=1$ system~\cite{Watanabe2006}. 
We note that the direction of node in d-wave pairing, 
i.e., $d_{x^2-y^2}$  or $d_{xy}$ symmetry, depends on the details of each system 
such as the shape of Fermi surface and the lattice structure. 
In principle, spin-triplet pairing, such as p-wave pairing, is not excluded 
because it is also included in a manifold of the intersite pairings.

\section{Comparison with experiments}

\subsection{$\beta$-YbAlB$_4$}

$\beta$-YbAlB$_4$ is a heavy fermion metal with intermediate valence, 
which is Yb$^{+2.75}$ at $T=20$~K~\cite{Okawa}.  
The characteristic energy scale corresponding to the so-called 
Kondo temperature is estimated to be $T_{\rm K}\sim 200$~K, which has  
a large c-f hybridization~\cite{Nakatsuji}.   
The uniform magnetic susceptibility and specific heat coefficient show 
a non-Fermi liquid behavior  
$\chi\sim T^{-0.5}$ and $C/T\sim -\log{T}$ for $0.3 \le T \le 2$~K, respectively, 
giving rise to large Wilson ratio $R_{\rm W}\approx 25$ 
at $T=0.4$~K~\cite{Nakatsuji,Matsumoto_JPCS2011}. 
A $T$-linear resistivity emerges in a wide temperature range for $1~{\rm K} \le T \le 4$~K and 
$\rho\sim T^{1.5}$ for $T < 1$~K~\cite{Nakatsuji}. 
These unconventional criticality summarized in Table~\ref{tb:exp} 
are well explained by the quantum valence criticality 
discussed in \S2.3 (see Table~\ref{tb:VQCP}). 
We note that the nuclear spin-lattice relaxation rate 
$(T_{1}T)^{-1}\sim T^{-\alpha}$ with $0.5 \lsim \alpha \lsim 0.7$ is predicted  
by the theory of quantum valence criticality, and 
the measurement of $(T_{1}T)^{-1}$ is highly desirable. 

\begin{table}
\begin{center}
\begin{tabular}{lcccccc} \hline
 & {$\rho$} & {$C/T$} & {$\chi$}  & {$(T_{1}T)^{-1}$} & $R_{\rm W}$ & reference
\\ \hline
YbRh$_2$Si$_2$ & {$T$} & {$-\log{T}$} &  $T^{-0.6}$ & $T^{-0.5}$ & 17.5
& \cite{Trovarelli,Gegenwart2007,Custers,Gegenwart2005,Ishida}
\\  
$\beta$-YbAlB$_4$ & {${T^{1.5} \to T \ }^{*}$} & {$-\log{T}$} & {$T^{-0.5}$} & ** & 
25 & \cite{Nakatsuji,Matsumoto2010,Matsumoto2011,Matsumoto_JPCS2011} 
\\
\hline
\end{tabular}
\end{center}
\caption{
Physical quantities of 
resistivity $\rho$ ($^*$~$\rho \sim T^{1.5}$ for $T<1$~K, $\rho \sim T$ for $T>1$~K), 
specific heat coefficient $C/T$, uniform magnetic susceptibility $\chi(T)$, 
nuclear spin-lattice relaxation rate $(T_1T)^{-1}$ 
i$^{**}$~experiment is desired), and Wilson ratio $R_{\rm W}$ at low temperatures in YbRh$_2$Si$_2$ and $\beta$-YbAlB$_4$. 
}
\label{tb:exp}
\end{table}

\subsection{YbRh$_2$Si$_2$} 

In YbRh$_2$Si$_2$~\cite{Trovarelli,Gegenwart2007} and 
YbRh$_2$(Si$_{0.95}$Ge$_{0.05}$)$_2$~\cite{Custers,Gegenwart2005}, 
unconventional criticality is observed as  
uniform magnetic susceptibility  
$\chi(T)\sim T^{-0.6}$, specific-heat coefficient $C/T\sim-\log{T}$ 
for $0.3 \le T \le 10$~K~\cite{Custers},  
giving rise to large Wilson ratio $R_{\rm W}\approx 17.5$ 
at $T=90$~mK~\cite{Gegenwart2005},   
$T$-linear resistivity for $20~{\rm mK}\le T \le 10$~K~\cite{Custers}, 
and spin-lattice relaxation rate 
$(T_{1}T)^{-1}\sim T^{-0.5}$ for $50~{\rm mK} \le T \le 1$~K~\cite{Ishida}. 
These are summarized in Table~\ref{tb:exp} and 
are well explained by the quantum valence criticality (see Table~\ref{tb:VQCP}). 
 
At $T=300$~K, the valence of Yb is observed as $\nu=+2.90$~\cite{Knebel2006}. 
This strongly suggests that when $T$ decreases, the Yb valence further decreases 
to be of the order of $\nu\sim +2.8$ near $T=0$~K. 
This implies that YbRh$_2$Si$_2$ is an intermediate valence material~\cite{Sekiyama}, 
similar to $\beta$-YbAlB$_4$~\cite{Okawa}. 
Recent angle-resolved photoemission measurement at $T=15$~K has revealed that 
the c-f hybridized band exists near the Fermi level in YbRh$_2$Si$_2$~\cite{Yasui}. 
This clearly indicates that the c-f hybridization exists at low temperatures 
in YbRh$_2$Si$_2$ for $H=0$. 

\begin{figure}
\begin{center}
\includegraphics[width=150mm]{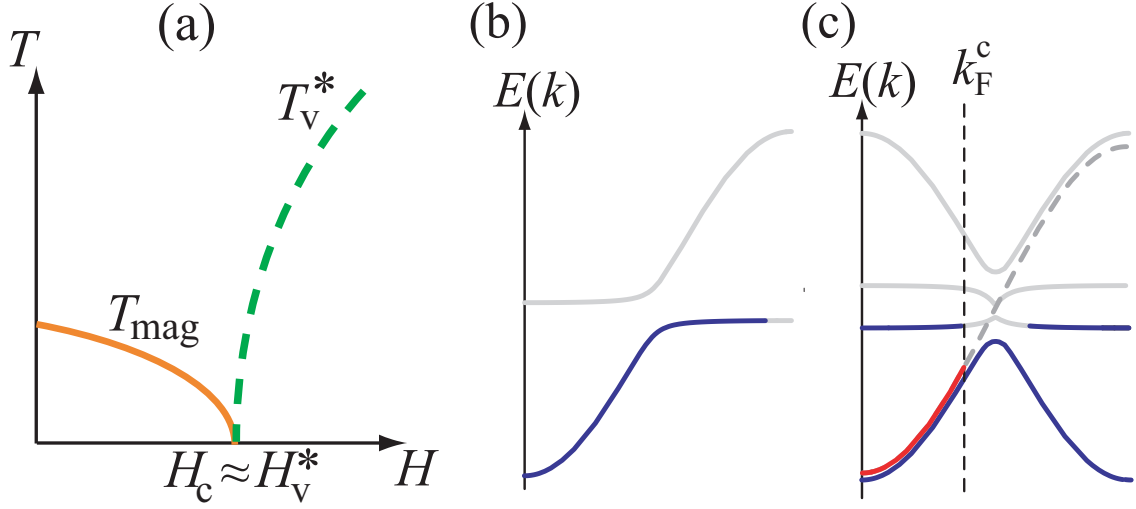}
\end{center}
\caption{(color online) 
(a) Schematic $T$-$H$ phase diagram in the case of small to moderate strength 
of c-f hybridization in ${\cal H}_{\rm PAM}$  (see text).
$T_{\rm mag}$ is the magnetic transition temperature and $T_{\rm v}^{*}$ is valence crossover 
temperature. 
Energy bands for (b) paramagnetic metal phase 
and (c) AF-ordered phase in ${\cal H}_{\rm PAM}$. 
In (c), the shaded dashed line represents energy band for conduction electrons. 
$k_{\rm F}^{\rm c}$ represents the small Fermi surface for only conduction electrons 
and the solid dashed line is a guide for the eyes (see text). 
}
\label{fig:TmagTv}
\end{figure}

In the $T$-$H$ phase diagram of YbRh$_2$Si$_2$, a characteristic temperature 
$T^{*}(H)$ appears, as illustrated by a dashed line in Fig.~\ref{fig:TmagTv}(a),  
where physical quantities show crossover behavior~\cite{Gegenwart2007}. 
The origin of $T^{*}(H)$ is considered to be closely related to the origin of 
unconventional criticality in YbRh$_2$Si$_2$. 
Here, we note that a field-induced QCEP of the FOVT can explain the emergence of 
the $T^{*}(H)$ line. 
Recently, magnetic field dependence of the QCEP has been theoretically studied on the basis of 
${\cal H}_{\rm PAM}$~(\ref{eq:PAM})~\cite{Watanabe2009,WTMF2008}. 
It has revealed that the QCEP extends to the smaller-$U_{\rm fc}$ region 
in Fig.~\ref{fig:TEfUfc} as magnetic field increases, 
which makes a sharp valence crossover line $T_{\rm v}^{*}(H)$ appear in the $T$-$H$ phase 
diagram, as shown in Fig.~\ref{fig:TmagTv}(a).

Recently, interplay of the QCEP of the FOVT and magnetism has been 
theoretically studied in ${\cal H}_{\rm PAM}$~\cite{WM2010,Watanabe2011}. 
For a realistic strength of c-f hybridization $|V_{\bf k}|$ in eq.~(\ref{eq:PAM}),  
a coincidence of magnetic transition temperature $T_{\rm mag}$ and sharp valence crossover 
temperature $T_{\rm v}^{*}$ occurs, as shown in Fig.~\ref{fig:TmagTv}(a)~\cite{WM2010}. 
When the hybridization decreases, $H_{\rm c}$ tends to be larger than $H_{\rm v}^{*}$ 
in Fig.~\ref{fig:TmagTv}(a). Then, enhanced valence fluctuations at $H=H_{\rm v}^{*}$ 
suppress the magnetic order, which still makes the coincidence of 
$H_{\rm c}\approx H_{\rm v}^{*}$. 
Namely, for realistic values of the c-f hybridization $|V_{\bf k}|$ 
ranging from the rather small to moderate strength, 
the coincidence of $T_{\rm mag}$ and $T_{\rm v}^{*}$ occurs at $T=0$~K.
This result well explains the $T$-$H$ phase diagram of YbRh$_2$Si$_2$. 

We also note that the $T$-$H$ phase diagram shown in Fig.~3(a) was actually realized in YbAuCu$_4$, where Cu-NQR frequency change was detected at $T_{\rm v}^{*}(H)$ clearly indicating Yb-valence crossover~\cite{Wada}. The phase diagram is very similar to the $T$-$H$ phase diagram in YbRh$_2$Si$_2$~\cite{wataPSS2010}. 
 
When we set a larger $|V_{\bf k}|$ in ${\cal H}_{\rm PAM}$, 
the magnetically ordered phase shrinks to be  
$H_{\rm c}<H_{\rm v}^{*}$ in Fig.~\ref{fig:TmagTv}(a). 
Namely, the magnetic quantum critical point (QCP) and 
the valence crossover point are separated. 
On the other hand, 
when we set very small c-f hybridization, $H_{\rm c}$ exceeds $H_{\rm v}^{*}$ to be 
$H_{\rm v}^{*}<H_{\rm c}$.  Namely, crossing of $T_{\rm mag}$ and $T_{\rm v}^{*}$ 
occurs. 

In the paramagnetic phase in YbRh$_2$Si$_2$, it was demonstrated that a tiny change in Yb valence, about 0.03, can reproduce the measured drastic change in the Hall coefficient by the band-structure calculation~\cite{Norman}. Namely, this suggests that the measured Hall-coefficient change at $T^{*}(H)$ in the $T$-$H$ phase diagram of YbRh$_2$Si$_2$ can be explained by the Yb-valence crossover. In ref.~\cite{Norman}, it was pointed out that a slight shift of the f level causes large change in the conductivities because of large density of states for 4f electrons. Here, it is noted that band-structure calculations for YbRh$_2$Si$_2$ have been performed by several groups~\cite{Knebel,Jeong,Rouke} and a renormalized-band approach has been also discussed~\cite{Friedemann_2010}. Although these methods do not seem to show complete agreements about detailed band structures and Fermi-surface shapes, they at least show large density of states for 4f electrons. Hence, it seems important to examine the possibility of whether a slight shift of the large density of states causes the drastic change of the Hall coefficient under a magnetic field within the band-structure calculations. 

Although the Kondo breakdown scenario has been intensively discussed to explain the field dependence of the Hall coefficient~\cite{Friedemann_2010b}, it would not be clear enough how it is  reconciled with the experimental fact that large effective mass ${\lim_{T\to 0}}C/T\sim 1.7$~J/molK$^2$ is observed at $H=0$~\cite{Krelner} (see also \S4). Here, we focus on the quantum phase transition between the AF- and paramagnetic-metal phases (see $H\sim H_{\rm c}$ in Fig.~\ref{fig:TmagTv}(a)) to address this issue from the viewpoint of the Yb-valence crossover.

In the paramagnetic metal phase, a large Fermi surface 
which includes an f-electron number is realized, as shown in Fig.~\ref{fig:TmagTv}(b).  
In the antiferromagnetic (AF) order phase, the hybridized bands are folded, 
as shown in Fig.~\ref{fig:TmagTv}(c). 
An important point here is that because the lowest folded band with 
one electron number per site shrinks down from the Fermi level, 
the folded band at the Fermi level has an electron number 
which is the same as that for conduction electrons. 
Namely, the Fermi surface in the folded band is the same as the small Fermi surface 
only made from the conduction electrons, as shown by $k_{\rm F}^{\rm c}$ 
in Fig.~\ref{fig:TmagTv}(c). 
Hence, a small to large Fermi surface change occurs at the AF to paramagnetic transition, 
even though c-f hybridization is always finite, i.e.,  
$\langle f^{\dagger}_{{\bf k}\sigma}c_{{\bf k}\sigma}\rangle\ne 0$.
Note that this is just a matter of number counting, which does not depend on the order 
of magnetic transition, i.e., the first-order or second order transition does not matter. 
Because heavy quasi particles contribute to the formation of the AF-ordered state, 
as shown in Fig.~\ref{fig:TmagTv}(c), 
mass enhancement occurs in the AF phase. 
This naturally explains the fact that heavy electron mass is observed 
as $C/T\sim 1.7$~J/molK$^2$ in the magnetically ordered phase 
in YbRh$_2$Si$_2$~\cite{Krelner}. 
These aspects cannot be understood in the so-called 
Kondo breakdown scenario~\cite{Si,Coleman}. 

To clarify the origin of unconventional criticality in YbRh$_2$Si$_2$, we propose 
to perform the Co-NQR measurement in the Co-doped sample: 
Yb(Co$_x$Rh$_{1-x}$)$_2$Si$_2$~\cite{Friedemann}. 
If the Co-NQR frequency $\nu_{\rm Q}$ changes at the characteristic temperature $T^{*}(H)$ 
in the $T$-$H$ phase diagram, it will be evidence 
that Yb-valence crossover occurs at $T^{*}(H)$ as observed in YbAuCu$_4$~\cite{Wada}. 
Measurement of this is highly desirable.

\subsection{YbCu$_{5-x}$Al$_{x}$}
\begin{figure}
\begin{center}
\includegraphics[width=60mm]{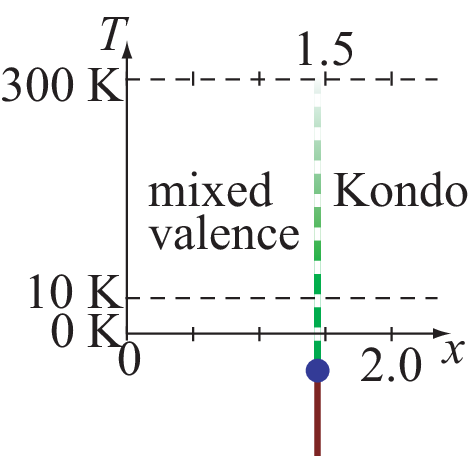}
\end{center}
\caption{(color online) 
Schematic phase diagram of YbCu$_{5-x}$Al$_x$~\cite{Bauer1997,Yamamoto,Yamaoka}. 
Near $x=1.5$ sharp Yb-valence crossover occurs between the Kondo regime and 
mixed valence regime at low temperatures $T\lsim 10$~K. 
}
\label{fig:YbCuAl}
\end{figure}

In the heavy fermion metal YbCu$_{5-x}$Al$_{x}$, 
at $T=300$~K, a gradual decrease in the Yb valence occurs as $x$ decreases from 2.0 to 0.0.
However, at $T=10$~K, a sharp decrease in the Yb valence occurs near $x=1.5$~\cite{Bauer1997}, 
which has been also supported by recent detailed measurements of the Yb valence~\cite{Yamamoto,Yamaoka}.  
These results indicate that a sharp Yb-valence crossover occurs near $x=1.5$ 
with strong Yb-valence fluctuations.  
The schematic $T$-$x$ phase diagram of YbCu$_{5-x}$Al$_{x}$ is shown in Fig~\ref{fig:YbCuAl}. 
At $x=1.5$, as $T$ decreases, logarithmic divergence in the specific-heat coefficient 
$C/T\sim -\log{T}$ 
and the power-law divergence in the uniform magnetic susceptibility $\chi\sim T^{-2/3}$ 
were observed in 1997 by E. Bauer {\it et al}~\cite{Bauer1997}, which well agree with 
the quantum valence criticality (see Table~\ref{tb:VQCP}).  

The shape of $T_{\rm v}^{*}(g)$, i.e.,  the control parameter dependence of the valence crossover temperature with $g$ being the control parameter such as a magnetic field, pressure and chemical substitution, depends on where the material is located in the phase diagram in Fig.~1. Detailed comparison of $T_{\rm v}^{*}(g)$ between the theory and experiments will be interesting future studies. 

\subsection{CeIrIn$_5$}

CeIrIn$_5$ is a heavy fermion metal which exhibits superconductivity at 
$T_{\rm c}\approx 0.4$~K~\cite{Petrovic}. 
Interestingly, 
by applying pressure, spin fluctuations are quickly suppressed, while  
superconducting transition temperature $T_{\rm c}(P)$ increases and 
has a maximum around $P\sim 2.5$~GPa, 
which suggests that the origin of superconductivity may be different from ordinary 
spin-fluctuation mechanism~\cite{Kawasaki}. 
Recently, M. Yashima and Y. Kitaoka have observed that 
In-NQR frequency $\nu_{\rm Q}$ starts to change sharply 
around $P\sim 2.2$~GPa 
(namely, $\nu_{\rm Q}-P$ relation changes its slope for $d\nu_{\rm Q}/dP$ to take a minimum as a function of $P$)~\cite{Yashima}, 
as observed in CeCu$_2$Si$_2$ at $P \sim 4$~GPa~\cite{Fujiwara}. 
This indicates that Ce-valence crossover sharply occurs around the pressure of 
$T_{\rm c}(P)$ maximum, strongly indicating that superconductivity is caused by 
the Ce-valence fluctuation. 

In the $T$-$H$ phase diagram, a first-order-transition or sharp-crossover line 
$T^*(H)$ appears~\cite{Takeuchi,Kim,Parm,Capan2004,Capan2009}, 
which is most likely to be FOVT or 
valence crossover~\cite{Watanabe2009}. 
Interestingly, quantum critical phenomena appear near $H=H^*$ where the $T^*(H)$ line 
touches $T=0$~K. 
As $H$ increases to approach $H^*$, convex curve in the $T^{1.5}$ plot of $\rho(T)$ 
emerges, indicating that $T$-linear resistivity appears near $H=H^*=28$~T~\cite{Capan2004}.  
Residual resistivity $\rho_{0}$ is enhanced around $H=H^*$~\cite{Capan2009}. 
Logarithmic divergence in specific heat $C/T\sim -\log{T}$ appears near $H=H^{*}$~\cite{Kim}. 
We pointed out that field dependence of the QCEP in ${\cal H}_{\rm PAM}$  
well explains the emergence of the $T^{*}(H)$ line in the $T$-$H$ phase diagram 
of CeIrIn$_5$~\cite{Watanabe2009,WTMF2008}. 

We here note that quantum valence criticality shown in Table~\ref{tb:VQCP} explains 
the measured criticality near $H=H^*$. 
Since the QCEP of the FOVT is induced by applying 
the magnetic field~\cite{Watanabe2009,WTMF2008}, 
quantum valence criticality 
emerges in high-field region, which is consistent with the observations near 
$H\sim H^*=28$~T in CeIrIn$_5$~\cite{Watanabe2009}. 
As shown in refs.~\cite{OnishiM,Watanabe2006}, 
the superconducting phase appears robustly even in the region rather far from the QCEP 
in the phase diagram of ${\cal H}_{\rm PAM}$ 
(see Fig.~\ref{fig:TEfUfc}). 
This is also consistent with the recent discovery by Yashima {\it et al},  
that at $H=0$, CeIrIn$_5$ seems to not be very close to the QCEP, 
but located in a gradual Ce-valence crossover region. 
Hence, quantum valence criticality based on ${\cal H}_{\rm PAM}$ gives a unified explanation 
for the measured anomalies in CeIrIn$_5$. 

These observations indicate that a new viewpoint of the closeness to the QCEP of the FOVT 
is indispensable for comprehensive understanding of Ce115 systems, 
in addition to the conventional picture of the so-called ``Doniach's phase diagram". 
Indeed, it has been shown theoretically 
that interplay of the QCEP of the FOVT and magnetism  
gives a unified explanation for several anomalies such as drastic change of Fermi surface 
and non-Fermi liquid transport observed in CeRhIn$_5$ under pressure~\cite{Watanabe2009,WM2010}. 
Namely, the results discussed for Fig.~\ref{fig:TmagTv}(a) in \S3.2 
basically hold even when the magnetic field is replaced by pressure 
as a control parameter~\cite{WM2010}. 
For details, readers can refer to refs.~\cite{WM2010,Watanabe2011}

\subsection{Discussion}

In CeCu$_2$Ge$_2$, X-ray absorption measurement under pressure reported 
that the lattice constant 
shows a sudden jump at $P=P_{\rm V}\sim 15$~GPa, 
where the superconducting transition temperature has a maximum~\cite{Onodera}. 
However, reexamination of the X-ray measurement in CeCu$_2$Ge$_2$ has observed 
a smooth decrease in the lattice constant under pressure but no jump at 
$P=P_{\rm v}$~\cite{TCKobayashi}. 
A smooth decrease in the lattice constant around $P\sim P_{\rm v}\sim 4.5$~GPa 
has been also observed in CeCu$_2$Si$_2$~\cite{TCKobayashi}.
We note that a sharp Cu-NQR-frequency change has been observed at $P\sim P_{\rm v}$ 
even without a jump in the lattice constant in CeCu$_2$Si$_2$~\cite{Fujiwara}. 
This indicates that NQR is more sensitive than the 
lattice-constant measurement, which can detect a tiny change of the electronic state 
due to the Ce- and Yb-valence change with the highest sensitivity. 

Here, let us remind the fact that even in Ce metal~\cite{Ce,Wohllleben} and 
YbInCu$_4$~\cite{Felner,Sarrao}, which show 
isostructural FOVT, the valence change is only about $0.1$. 
In Ce metal, isostructural first-order valence transition is known to occur, 
well known as $\gamma$-$\alpha$ transition~\cite{Ce}. 
At $T=300$~K, the Ce valence 
shows a discontinuous jump from Ce$^{+3.03}$ to Ce$^{+3.14}$ 
at the $\gamma$-$\alpha$ transition, as pressure increases. 
The lattice constant also shows a discontinuous jump at the $\gamma$-$\alpha$ transition, 
about 16$\%$ shrinkage of the unit-cell volume. 
In YbInCu$_4$, 
isostructural FOVT is also known to occur~\cite{Felner}. 
As temperature decreases, the Yb valence shows a 
discontinuous change from Yb$^{+2.97}$ to Yb$^{+2.84}$ with a volume expansion, i.e.,  
0.5$\%$ increase in the unit-cell volume, at $T_{\rm v}=42$~K~\cite{Cornelious,Dallera,Matsuda2,Suga}.  
Hence, in the valence crossover regime where 
most of Ce- and Yb-based heavy fermion systems are considered to be located, 
the valence change is expected to be in the order of $0.01$, 
even when sharp valence crossover occurs (see Fig.~\ref{fig:TEfUfc}).

Note that the volume change in YbInCu$_4$ 
is much smaller than that in Ce metal, 
although Ce metal and YbInCu$_4$ both show a jump in Ce and Yb valence of 
about $\sim 0.1$ at the FOVT, respectively.
There is a tendency that for volume change in Ce- and Yb-based compounds to be 
much smaller than Ce (or Yb) mono metal.
Hence, it should be kept in mind that even in the cases where the lattice constant merely shows a 
monotonic change as control parameters, such as pressure and a magnetic field, are changed, 
there is a possibility that NQR frequency change is detectable. 

Recently, the pressure dependence of Ce valence in CeCu$_2$Si$_2$ has been observed by  
X-ray $L_{\rm III}$ edge absorption spectra~\cite{Rueff}.  
Although the data have error bars of the order $\sim 0.01$ and 
discretely measured pressure data points do not seem to be 
enough to see whether sharp valence crossover in the order of $\sim 0.01$ 
occurs around $P_{\rm v}\sim 4.5$~GPa, a possibility of 
monotonic Ce-valence crossover is not excluded. 
Since the measurement has been performed at $T=14$~K, this result may suggest  
that the QCEP of the FOVT is located at much lower temperature, 
as shown in Fig.~\ref{fig:TPD}(c).  

In our studies, we have neglected effects of excited state(s) of 
crystalline-electric-filed (CEF) levels. For the moment, it is not clear 
whether those CEF states give an essential effect for realistic 
CEF states while an effect of orbital transfer between CEF levels 
(with a model with hybridization with conduction electrons) 
was discussed by Hattori~\cite{Hattori}. 
In any case, such effects certainly deserve more investigations. 

\section{Discussion about other theoretical proposals}

In this paper, we have discussed that fundamental properties of the outstanding materials such as YbRh$_2$Si$_2$ are explained coherently from the viewpoint of quantum valence criticality. So far, to understand the unconventional criticality, several scenarios such as the local criticality theory~\cite{Si,Coleman,Si_2006} and theory of tricritical point (TCP)~\cite{Misawa} were proposed. Here, we discuss them within the scope of the present paper. 

The local criticality theory did not fully account for the unconventional criticality in YbRh$_2$Si$_2$ shown in Table~2. In ref.~[9], it was shown that the magnetic susceptibility behaves as $\chi(T)=1/(\Theta +\tilde{A}T^{\alpha})$. However, the measured value of $\alpha\sim 0.6$ was not shown explicitly by the theory~\cite{Si,Si_2006}. The theory concluded that $1/T_{1}={\rm constant}$, which does not reproduce the measurement $(T_{1}T)^{-1}\sim T^{-0.5}$ for low temperatures $(50~{\rm mK}<T<1~{\rm K})$~\cite{Ishida}. Furthermore, the $T$-linear resistivity and logarithmic-$T$ dependence of $C/T$ were not shown by the theory itself~\cite{Si,Si_2006}. In ref.~\cite{Coleman}, a scenario of Kondo breakdown was just proposed, but the temperature dependences in Table~2 were not shown theoretically. 
%
In ref.~\cite{Coleman}, the transition between the conventional SDW-type AF metal with reconstructed Fermi surfaces around ``hot spots" and the paramagnetic metal (PM) with the large Fermi surface is argued. We note here that the ``small"-
to-large Fermi surface change can occur at the SDW-type AF-PM transition in the ground state of the periodic Anderson model for realistic parameters for heavy electron systems~\cite{WM2010,WO2007} (see Fig.~\ref{fig:TmagTv}(c) in \S3.2). Namely, Fermi surface changes from the ``small" Fermi surface, which is the same as that of the non-hybridized conduction band, realized in the SDW-type AF phase to the large Fermi surface in the PM phase.
Furthermore, as noted in \S3.2, it is not clear how measured large $C/T\sim 1.7$~J/molK$^2$~\cite{Krelner} is reconciled with switching off the c-f hybridization in the magnetically-ordered phase. Although the concept of ``dynamical Kondo screening" was proposed  to explain the measured  large mass~\cite{Si_2006}, it seems necessary to show explicitly how large mass can appear in their model, i.e., the Kondo lattice model with the RKKY interaction model, for reliable argument.
It is also noted that in the most of Ce- and Yb-based materials, intermediate valences of Ce and Yb are realized near $T=0$~K, implying the existence of valence fluctuations at low temperatures.

The TCP theory shows at low temperatures the uniform magnetic susceptibility $\chi(T)\sim T^{-0.75}$, the nuclear spin-lattice relaxation rate $(T_{1}T)^{-1}\sim  -D_{0}\log{T}+D_{Q}T^{-0.75}$, the resistivity $\rho(T)\sim T^{1.5}$, and the Sommerfeld  coefficient $C/T\sim {\rm constat}-T^{1/2}$ ($T<1$~K) $-\log{T}$ ($T>1$~K)~\cite{Misawa}. Although it may be possible to interpret that some agreements with the measurements can be seen for some temperature regions, these results are not exactly the same as the measurements shown in Table~2. In particular, it should be noted that the TCP theory did not show the $T$-linear resistivity at low temperatures which is the prominent feature in YbRh$_2$Si$_2$. Furthermore, it is not clear why the characteristic temperature $T^{*}(H)$ appears in the $T$-$H$ phase diagram of YbRh$_2$Si$_2$, which is  closely related to the origin of the unconventional criticality. 

It is remarked that by the interplay of the magnetically ordered temperature and the sharp-valence crossover temperature as a function of control parameters such as pressure,  magnetic field, or chemical substitution, the first-order magnetic transition can be caused, giving rise to the TCP, as shown in ref.~\cite{WM2010,Watanabe2011}. Search for such a material is left for future experimental studies.

\section{Summary and Perspective}

By constructing a mode-coupling theory for critical valence fluctuations 
taking account of strong on-site Coulomb repulsion in the periodic Anderson model 
with inter-orbital Coulomb repulsion, we have shown that a new class of universality 
emerges near the QCEP of the FOVT. 
The quantum valence criticality 
gives a unified explanation for unconventional criticality observed in 
heavy fermion metals such as 
$\beta$-YbAlB$_4$, YbRh$_2$Si$_2$, YbCu$_{5-x}$Al$_{x}$, and CeIrIn$_5$. 

The key origin of emergence of unconventional criticality is clarified to be 
due to the extremely dispersionless critical valence fluctuation mode caused by 
the strong on-site Coulomb repulsion $U$ for f electrons. 
Observation of the locality of the critical valence-fluctuation mode is 
an important future issue for revealing direct evidence of 
the origin of the unconventional criticality. 
Resonant X-ray scattering offers a direct observation of $q$ dependence of the valence susceptibility  $\chi_{\rm v}^{\rm R}(q,\omega)$. 
As discussed in \S2, an almost dispersionless critical valence fluctuation mode is expected to 
be observed near the QCEP of the FOVT. 
To detect such a tiny width of dispersion, 
high resolution of resonant X-ray scattering measurement is necessary.  
Since dynamical f-spin susceptibility $\chi^{\rm R}(q,\omega)$ is considered to 
have a common structure with  
the valence susceptibility $\chi_{\rm v}^{\rm R}(q,\omega)$~\cite{VQCP2010}, 
neutron measurements for materials 
located in the vicinity of the QCEP of the FOVT are also useful for 
examining the possibility of whether an almost $q$-independent mode appears near $q=0$.  

In this paper, we focused on Ce- and Yb-based heavy fermion systems. 
Even in the other materials, there remains a possibility that 
quantum valence criticality emerges, since 
quantum valence criticality arises from fluctuations of charge transfer between orbitals 
in itinerant electron systems with strong local electron correlations, which 
are quite common. Hence, it seems to have wider relevance. 
The experimental challenge to uncover new universality class in correlated electron systems 
is an interesting future issue. 

\section*{Acknowledgement}

This work is supported by a Grant-in-Aid for Scientific Research on Innovative Areas ``Heavy Electrons" (No. 20102006 and No. 20102008) from the Ministry of Education, Culture, Sports, Science, and Technology, Japan (MEXT). One of us (S.~W.) is supported by the exciting leading-edge research project ``TOKIMEKI" at Osaka University and a Grant-in-Aid for Scientific Research (C) (No. 24540378) from MEXT. One of us (K.~M.) is supported by a Grant-in-Aid for Specially Promoted Research (No. 20001004) from MEXT.

\end{document}